\documentclass{IEEEcsmag}

\usepackage[colorlinks,urlcolor=blue,linkcolor=blue,citecolor=blue]{hyperref}

\usepackage{upmath}

\begin{document}

\title{Advantages of maintaining a multi-task project-specific bot: an experience report}

\author{Théo Zimmermann}
\affil{Inria, Université Paris Cité, CNRS, IRIF, Paris, France}

\author{Julien Coolen}
\affil{Université Paris Cité, Paris, France}

\author{Jason Gross}
\affil{MIRI, USA}

\author{Pierre-Marie Pédrot}
\affil{Inria and LS2N, Nantes, France}

\author{Gaëtan Gilbert}
\affil{Inria and LS2N, Nantes, France}

\begin{abstract}
Bots are becoming a popular method for automating basic everyday tasks in many software projects. This is true in particular because of the availability of many off-the-shelf task-specific bots that teams can quickly adopt (which are sometimes completed with additional task-specific custom bots). Based on our experience in the Coq project, where we have developed and maintained a multi-task project-specific bot, we argue that this alternative approach to project automation should receive more attention because it strikes a good balance between productivity and adaptibility. In this article, we describe the kind of automation that our bot implements, what advantages we have gained by maintaining a project-specific bot, and the technology and architecture choices that have made it possible. We draw conclusions that should generalize to other medium-sized software teams willing to invest in project automation without disrupting their workflows.
\end{abstract}

\maketitle

\chapterinitial{On collaborative coding platforms} like GitHub, bots are commonplace nowadays~\cite{wessel2018power}, as it has become easy and encouraged to add new bots to one's projects.
This is great for small teams or single developers wanting to quickly speed up their project's adoption of best practices. But for larger teams with well-established processes, having to adapt to rigid workflows of preexisting bots can be disruptive.

Previous research has already shown that task-oriented GitHub bots can cause friction by lacking social context or disrupting developers' workflows~\cite{brown_sorry_2019}. Wessel \emph{et al.}~\cite{wessel2022bots} have proposed the promising concept of a meta-bot (aggregating and summarizing information coming from several bots) to alleviate these issues. For several years, we have explored another strategy, that shares some characteristics with a meta-bot: relying on a multi-task, project-specific bot, directly developed and maintained by the project team.
The bot works hand-in-hand with developers, helping them by automating everything that is repetitive and easily automatable, without requiring changes to their workflow. For medium to large teams, this can be a reasonable investment to make, that will be largely compensated by the returns.

We have adopted this strategy for the maintenance of the Coq proof assistant~\cite{the_coq_development_team_2021_5704840}, a medium-sized open-source software system, managed by a core team of about 10 developers, an extended maintainer team of about 30 people, and hundreds of contributors.
In this article, we describe how we have developed such a bot and the kind of tasks it helps with. From our experience, we conclude that there are many benefits maintaining a multi-task project-specific bot, and we draw lessons that could help other software teams follow a similar approach.

\section{BOT INTERACTIONS IN THE PULL REQUEST LIFECYCLE}

The Coq bot interacts with Coq contributors at several stages of the pull request (PR) lifecycle.

\subsection{Triggering CI and reporting results}

When a pull request is opened (or updated), the bot takes care of triggering and reporting the results of Continuous Integration (CI) testing.

Even if it is hosted on GitHub, the Coq project relies mainly on GitLab CI, because it is one of the rare CI platforms that can stand the extensive use of CI by the project (where each PR will trigger the build of dozens of reverse dependencies for compatibility testing)~\cite{zimmermann:tel-02451322}.

GitLab CI is marketed as being a possible CI solution for GitHub, but support is actually limited because it does not handle PRs coming from GitHub forks.
Therefore, the bot pushes and updates branches on the GitLab mirror for any opened or updated PR on the GitHub repository.

Even though GitLab supports reporting CI results back to GitHub, the bot handles this as well. This has several advantages. First, the bot will not report only the overall pipeline status but also any failed jobs (but not successful jobs, except in limited circumstances, because there are too many of them).

Second, to avoid giving false confidence on the impact of a PR when the PR branch seriously lags behind the base, the bot automatically creates merge commits between the PR head and the head of the base branch (this feature is inspired by the behavior of Travis CI).
Controlling the status report to GitHub was essential to implement this solution, since the bot can map from the tested merge commit to the origin GitHub commit.

When the Checks tab was introduced~\cite{github_checks_announcement},
we started relying on it to report CI log summaries directly on GitHub. Because the bot is project-specific, we can automatically search for errors in CI logs (based on knowledge of their expected shape) to ensure that we display them.

Following suggestions from PR reviewers, CI reports also include direct links to HTML documentation CI artifacts to ease previewing of documentation modifications.

While the bot is project-specific, and this feature is customized to be particularly suited for the Coq project, its core is of general interest and has been used beyond the repositories maintained by the Coq team.
Most of the other users are from the Coq ecosystem (e.g., the MathComp library), but some come from outside (e.g., the saltstack-formulas organization's hundreds of repositories).

\subsection{Triggering a test case reduction procedure}

The control over the CI reporting mechanism has allowed us to plug in an advanced feature to automatically minimize compatibility issues detected on reverse dependencies tested in Coq's CI. The bot automatically identifies such test failures and proposes to trigger the reduction process.
The users can do so by leaving a comment with the message ``@coqbot: ci minimize'' (or a more advanced variant). We do not describe this feature in detail here because it is the topic of another article~\cite{gross:hal-03586813}.

\subsection{Closing stale PRs}

PRs with merge conflicts with the base branch are automatically labeled with \texttt{needs: rebase}. To reduce the number of stale open PRs, the Coq team decided to introduce a policy to close PRs that had this label set for more than 30 days, after a warning and an additional 30-day grace period. This policy is enforced by the bot and is similar to what ``stale bots'' implement~\cite{wessel_should_2019}, but with a different criterion to determine that a PR is stale, since it relies on merge conflicts instead of the absence of any activity. While it means that some PRs are not considered as stale even if they have been inactive for a while, it also means that the required action to remove the stale status is more demanding than just posting a comment (it requires solving merge conflicts).

\subsection{Merging pull requests}

The Coq team has precise rules on when and how to merge a PR, in terms of labels, milestones, assignees, reviews, target branches, etc. Furthermore, a merge commit is required, with a message in a specific format and a PGP signature.
For several years, a merge script was available to check these requirements and apply the required formatting. However, it still represented a barrier to onboarding new maintainers (especially because of the requirement for signed merge commits).

We added support allowing maintainers to request the bot to merge a PR (by commenting ``@coqbot: merge now''). The bot then checks that all requirements are met and that the maintainer is an authorized maintainer before performing the merge (see \textbf{Figure \ref{fig:merge-backport}}). Internally, it relies on GitHub's merge button to produce a signed merge commit, but it checks many things that this merge button alone would not check and uses the expected formatting. If some conditions are not met, it responds with a comment explaining which ones.

Since it was introduced, this has been the dominant method for merging PRs, by all maintainers. Some new maintainers have never called the merge script (that is still available as an alternative, for now).
Implementing more advanced merging strategies with the bot (such as, after a final comment period, or after CI has completed successfully) is currently being discussed.

\subsection{Keeping track of the backporting process}

\begin{figure*}
\centerline{\includegraphics[width=34pc]{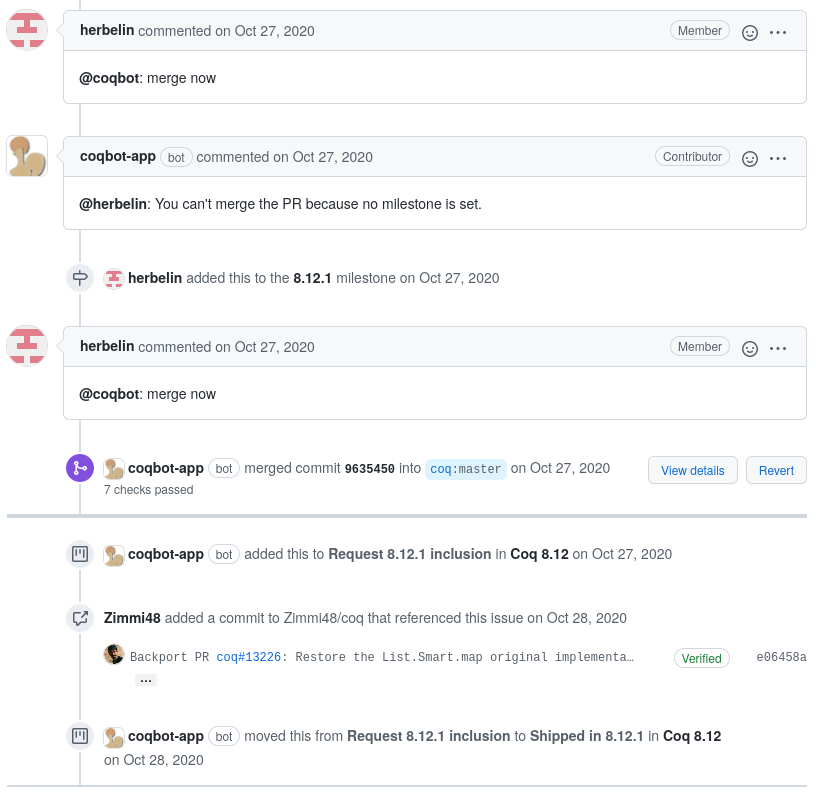}}
\caption{This screenshot demonstrates two features of the Coq bot. First, a maintainer uses the bot to merge a PR, but the bot reminds the maintainer that they have forgotten to set a milestone. After this is fixed, all the criteria for merging a PR are met, so the bot executes the command. Second, the bot analyzes the milestone of the merged PR to figure out if backporting was requested. In this case, it was, so it adds the PR to the appropriate ``Backport requested'' column of the RM backporting project. The RM then prepares the backport (on their fork), and when they push it to the release branch, the bot moves the PR to the corresponding ``Shipped'' column of the backporting project.}
\label{fig:merge-backport}
\end{figure*}

The Coq release management process is based on release branches that are controlled by a Release Manager (RM), which is a rotating position. PR authors and shepherds can signal that a PR should be backported by using an appropriate milestone, but ultimately, the decision is made by the RM.

The RM is helped by the bot which automatically tracks for which PRs backporting was requested (based on the milestone) and which PRs were already backported, in a dedicated GitHub project board (see Figure \ref{fig:merge-backport}).

The bot also handles backport rejections. The RM can reject a backport by removing a PR from the project board. In this case, the bot changes the milestone of the PR and posts a comment to inform PR stakeholders of the decision.

\section{ADVANTAGES OF A MULTI-TASK PROJECT-SPECIFIC BOT}

\subsection{Advantages of a project-specific bot}

As we have seen, many of the features implemented by the bot are similar to features proposed in off-the-shelf solutions, but differ in subtle ways (our CI reporting feature goes beyond what GitLab supports, our stale criterion differs from the one implemented in stale bots, the merging feature checks more than what would be possible with the GitHub button).

When an off-the-shelf solution is used, customization is limited to what is supported in configuration files. Going beyond requires getting hold of the code base, modifying it, deploying it, and maintaining it.

Instead of striving to adopt off-the-shelf solutions with their limited configurability, that would require less maintenance but make the project more dependent on external maintainers, and require the team to adapt its workflows, we inspire from existing solutions and implement them as we see fit, using our preexisting bot code base, which we have maintained for several years (since 2018). That being said, we do not forbid ourselves from actually installing off-the-shelf solutions if they do feel appropriate, or to test them before deciding what we would like to implement in our project-specific bot.

\subsection{Advantages of a multi-task bot}

The first advantage we gain by having a single bot that combines many features is to reduce the cognitive load for Coq contributors, who do not need to remember which bot has which feature and how it is triggered. This is in line with the strategy proposed by Wessel \emph{et al.}~\cite{wessel2022bots} to combine several (off-the-shelf) bots into a single meta-bot that provides a single interface to the wide range of their features.

Furthermore, features that may appear as independent without looking closely actually benefit from being implemented together. For instance, the CI reporting feature was modified to preserve information that would be useful to the test-case reduction feature. Similarly, tracking changes to release branches requires knowing how merge commits are formatted (and how to extract information from them). A new feature under testing to trigger benchmarks and report their results is plugged into the command mechanism (already used for the merging and test-case reduction features) and in the CI reporting mechanism (to replace the usual reporting by one that is more customized).

Finally, having all the project automation implemented into a single bot code base reduces the maintenance work significantly. Many components are actually reused across several features and fixing or evolving them can be done once and impact all the dependent features.

\section{TECHNOLOGY AND ARCHITECTURE}

Our bot is project-specific: it was built to assist the Coq team, and to evolve based on their feedback. To facilitate its evolution and the involvement of any Coq developer in the bot maintenance, we chose to rely on familiar technology, and to design an easy to understand and easy to extend architecture.

\subsection{Familiar technology}

A standard choice to develop GitHub bots is Probot~\cite{probot}, a Node.js framework for GitHub Apps. However, we decided to write the bot using OCaml~\cite{leroy:hal-00930213} instead.
Indeed, this is the programming language used to build Coq, which means that Coq developers are already familiar with it. This is also a strongly-typed language, thus it provides high confidence when introducing and refactoring code, which is something else that Coq developers are quite used to.

To maximize productivity, the bot depends on many OCaml libraries (to set up a web server, encode and decode JSON, etc.). This is standard software engineering practice, but this contrasts with the practice followed in the Coq codebase, where any new dependency is carefully evaluated to guarantee stability and facilitate distribution.

Among the dependencies, graphql-ppx~\cite{graphql-ppx} is used to interface with GitHub's GraphQL API.
This API enables querying for exactly the information we need, while reducing the number of requests, and providing more safety on the request correctness thanks to the typed GraphQL language and API.

\subsection{Straightforward and extensible architecture}

The bot is architectured around a library of reusable bot components, and an application of bot workflows. The bot components are reusable bricks that can be combined into different workflows, following trigger-action programming.

Trigger-action programming~\cite{huang2015supporting} is a programming model that has mostly been studied in the context of smart-home automation, with the idea of providing a programming framework and mental model that is accessible to anyone.
Famous trigger-action programming platforms are IFTTT and Zapier~\cite{rahmati2017ifttt}.
Interestingly, both provide GitHub integration (Zapier also provides GitLab integration), but their triggers and actions are not sufficiently advanced for our purposes.

Nowadays, another example of trigger-action programming is GitHub workflows, which are built by combining event triggers with prebuilt or custom ``GitHub Actions''. Many tasks bots perform can also be programmed using GitHub workflows~\cite{kinsman_how_2021} but with lower reactivity, because workflows need to start virtual machines to react to events.

Our bot components are divided into the three usual types of trigger-action programming~\cite{huang2015supporting}:
\begin{itemize}
\item \emph{Event triggers}: events that the bot listens to, by subscribing to GitHub / GitLab webhooks. For instance, a new comment is an event trigger that is reused in several workflows.
\item \emph{State triggers}: additional data needed to perform an action, obtained by querying web APIs. It does not need to exactly match a function from the API. For instance, a test that a user belongs to a given team is a state trigger.
\item \emph{Actions}: state-changing requests that are sent by the bot, acting as an agent on the platform. For instance, adding a label on an issue or PR is an action that is reused in several workflows.
\end{itemize}

Introducing new bot workflows is as easy as combining the various available components, or introducing new ones when needed. The use of GraphQL to interact with GitHub makes it easy to add state triggers or actions safely.

\subsection{Team involvement}

This architecture, the use of a language that the Coq team already masters, and of external libraries as often as needed, have helped the onboarding of new bot maintainers: the first author is the initial developer and maintainer of the Coq bot since 2018; the second author was an undergraduate summer intern in 2020 who significantly extended the bot with new features and helped complete the envisioned architecture; the three other authors are Coq developers who improved and extended the bot, sometimes with little help from the initial developer.

\section{LESSONS LEARNED}

While more and more projects adopt off-the-shelf bots to help them automate their everyday tasks, the experience of the Coq project of developing and maintaining a multi-task project-specific bot shows that this approach can be a successful alternative to boost developers' productivity while avoiding disruption of their pre-established workflows.

Of course, this approach requires some investment, so it is not suited for projects that are too small, but the required investment in development and maintenance is reasonable for medium-sized projects, especially when maintaining a single bot code base can facilitate the addition of many features, and when reusing familiar technology can enable everyone in the team to participate in the bot maintenance and evolution.

For projects from the OCaml ecosystem, we think that our library of bot components, although still experimental, could serve as a basis for other project-specific bots. In fact, a project-specific bot for another OCaml-based project (Usaba) is currently being developed by reusing our library, and its developer is already contributing changes back.

For projects in ecosystems that do not have such libraries, we think that creating similar bot component libraries would be useful to facilitate the application of our approach to projects of those ecosystems.

\section{ACKNOWLEDGMENT}

We thank Jean-Rémy Falleri and Thomas Degueule for providing feedback on early drafts of this paper, as well as the anonymous reviewers for their valuable feedback.

\bibliographystyle{ieeetr}
\bibliography{biblio}

\begin{thebibliography}{10}

\bibitem{wessel2018power}
M.~Wessel, B.~M. De~Souza, I.~Steinmacher, I.~S. Wiese, I.~Polato, A.~P.
  Chaves, and M.~A. Gerosa, ``The power of bots: Characterizing and
  understanding bots in {OSS} projects,'' {\em Proceedings of the ACM on
  Human-Computer Interaction}, vol.~2, no.~CSCW, pp.~1--19, 2018.

\bibitem{brown_sorry_2019}
C.~Brown and C.~Parnin, ``Sorry to bother you: Designing bots for effective
  recommendations,'' in {\em 2019 {IEEE}/{ACM} 1st {International} {Workshop}
  on {Bots} in {Software} {Engineering} ({BotSE})}, pp.~54--58, IEEE, May 2019.

\bibitem{wessel2022bots}
M.~Wessel, A.~Abdellatif, I.~Wiese, T.~Conte, E.~Shihab, M.~A. Gerosa, and
  I.~Steinmacher, ``Bots for pull requests: The good, the bad, and the
  promising,'' in {\em Proceedings of the 44th ACM/IEEE International
  Conference on Software Engineering (ICSE'22)}, vol.~26, p.~16, ACM/IEEE,
  2022.

\bibitem{the_coq_development_team_2021_5704840}
{The Coq Development Team}, ``The {Coq} proof assistant,'' Oct. 2021.

\bibitem{zimmermann:tel-02451322}
T.~Zimmermann, {\em {Challenges in the collaborative evolution of a proof
  language and its ecosystem}}.
\newblock {PhD} thesis, {Universit{\'e} de Paris}, Dec. 2019.

\bibitem{github_checks_announcement}
K.~McMinn, ``New checks {API} public beta,'' 2018.
\newblock
  \url{https://developer.github.com/changes/2018-05-07-new-checks-api-public-beta/}.

\bibitem{gross:hal-03586813}
J.~Gross, T.~Zimmermann, M.~Poddar-Agrawal, and A.~Chlipala, ``{Automatic
  Test-Case Reduction in Proof Assistants: A Case Study in Coq},'' in {\em
  Proceedings of the 13th Conference on Interactive Theorem Proving}, 2022.

\bibitem{wessel_should_2019}
M.~Wessel, I.~Steinmacher, I.~S. Wiese, and M.~A. Gerosa, ``Should {I} stale or
  should {I} close? {An} analysis of a bot that closes abandoned issues and
  pull requests,'' in {\em 2019 {IEEE}/{ACM} 1st {International} {Workshop} on
  {Bots} in {Software} {Engineering} ({BotSE})}, (Montreal, QC, Canada),
  pp.~38--42, IEEE, May 2019.

\bibitem{probot}
B.~Keepers and {Probot contributors}, ``Probot: A framework for building
  {GitHub Apps} to automate and improve your workflow,'' 2016--2021.
\newblock \url{https://github.com/probot/probot}.

\bibitem{leroy:hal-00930213}
X.~Leroy, D.~Doligez, A.~Frisch, J.~Garrigue, D.~R{\'e}my, and J.~Vouillon,
  ``{The OCaml system release 4.11: Documentation and user's manual},'' intern
  report, {Inria}, Aug. 2020.

\bibitem{graphql-ppx}
M.~Hallin, T.~Cichocinski, and J.~Frolich, ``graphql-ppx,'' 2017--2021.
\newblock \url{https://github.com/reasonml-community/graphql-ppx}.

\bibitem{huang2015supporting}
J.~Huang and M.~Cakmak, ``Supporting mental model accuracy in trigger-action
  programming,'' in {\em Proceedings of the 2015 International Joint Conference
  on Pervasive and Ubiquitous Computing}, pp.~215--225, ACM, ACM Press, 2015.

\bibitem{rahmati2017ifttt}
A.~Rahmati, E.~Fernandes, J.~Jung, and A.~Prakash, ``{IFTTT} vs. {Zapier}: A
  comparative study of trigger-action programming frameworks,'' {\em arXiv
  preprint arXiv:1709.02788}, 2017.

\bibitem{kinsman_how_2021}
T.~Kinsman, M.~Wessel, M.~A. Gerosa, and C.~Treude, ``How do software
  developers use {GitHub} {Actions} to automate their workflows?,'' {\em
  arXiv:2103.12224 [cs]}, Mar. 2021.
\newblock arXiv: 2103.12224.

\end{thebibliography}

\end{document}